\documentclass{ws-procs9x6-cpt22}
\begin{document}

\newcommand{\refeq}[1]{(\ref{#1})}
\def\etal {{\it et al.}}

\title{$\kappa$-deformed complex fields, (discrete) symmetries,\\ and charges}

\author{A.\ Bevilacqua$^1$}

\address{$^1$National Centre for Nuclear Research,\\
ul. Pasteura 7, 02-093 Warsaw, Poland}



\allowdisplaybreaks

\begin{abstract}
 
In what follows, we will briefly describe how to build a field theory of a complex scalar field in the $\kappa$-Minkowski spacetime. After introducing the action, we will shortly describe its properties under both continuous and deformed symmetry transformations. We will then describe how to compute the charges and describe their non-trivial properties due to $\kappa$-deformation. We will conclude with the experimental significance of the model, particularly in the context of decay probability differences between particles and antiparticles.
\end{abstract}

\bodymatter

\section{From non-commutative spacetime to the action}

What follows is a short preview of the works in Refs.\ \refcite{ref1}, \refcite{ref2}. For a more comprehensive and in depth introduction to this framework, see for example Ref.\ \refcite{ref3}.
The starting point of our considerations is given by Minkowski spacetime, which can in turn be defined in terms of non-commuting coordinates satisfying the $\mathfrak{an}(3)$ algebra defined as $[\hat{x}^0, \hat{\mathbf{x}}^i] = i\hat{\mathbf{x}}^i/\kappa$. Notice that $1/\kappa$ has dimensions of length\footnote{The formal limit $\kappa\rightarrow\infty$ reduces us to the canonical commutative Minkowski spacetime.}, and that these commutation relations are very different from the Moyal-type non-commutativity $[x^\mu,x^\nu]= i \theta^{\mu\nu}$. The most direct way to build easily manageable fields on this spacetime goes along the following lines. We first need plane waves, corresponding to the $AN(3)$ group elements $\hat{e}_k = \exp (i \mathbf{k}_i \hat{\mathbf{x}}^i) \exp (i k_0\hat{x}^0)$. Because of the group structure, addition of momenta is given by the group property, i.e., $\hat{e}_k \hat{e}_l = \hat{e}_{k\oplus l}$. In the same way, inverses of momenta are given by group inverses, i.e., $\hat{e}^{-1}_k = \hat{e}_{S(k)}$. We then employ a\footnote{There are many equivalent ways in which one could choose a suitable Weyl map. For a more in depth discussion, see Ref.\ \refcite{ref1}.} Weyl map $\mathcal{W}$, which sends group elements $\hat{e}_k$ to canonical plane waves $e_p := \exp(ip_\mu(k) x^\mu)$. The coordinates $x^\mu$ are now canonical coordinates, and the group structure is preserved through the introduction of a $\star$ product defined by $\mathcal{W}(\hat{e}_{k\oplus l}) = e_{p(k) \oplus q(l)}=: e_p \star e_q$. The $\star$ product is in general non-commutative. Because of this, considering the case of a complex scalar field, one has two possible choices of action, depending on whether one chooses the ordering $\phi\star \phi^\dag$ or $\phi^\dag \star \phi$ (and the same for the kinetic term, we are still considering a free action). The action we chose is
\begin{equation}
	S = \frac{1}{2} \int_{\mathbb{R}^4} d^4x [(\partial^\mu\phi)^\dag \star (\partial_\mu\phi) + (\partial^\mu\phi) \star (\partial_\mu\phi)^\dag - m^2 (\phi^\dag \star \phi + \phi\star \phi^\dag)].
	\label{aba:1}
\end{equation}
One can show that the field satisfies the Klein-Gordon equations, and that such a field can be described using the plane waves described above as
\begin{equation}
	\phi(x)=
	\int \frac{d^3p}{\sqrt{2\omega_p}} \xi(p) a_\mathbf{p}
	e^{-i(\omega_pt - \mathbf{p}\mathbf{x})}
	+
	\int \frac{d^3p^*}{\sqrt{2|\omega_p^*|}} \xi(p) b^\dag_{\mathbf{p}^*}
	e^{i(S(\omega_p)t - S(\mathbf{p})\mathbf{x})}.
	\label{aba:2}
\end{equation}
Notice the presence of the antipode in the conjugate plane wave\footnote{The ${}^*$ on the momenta of the second wave is related to a second copy of momentum space, for more details, see Ref.\ \refcite{ref1}.}. Because of this feature, and thanks to our choice of action, the CPT transformation of the field are the same as in the non-deformed case, i.e., T$\phi(t,\mathbf{x})$T${}^{-1}=\phi(-t,\mathbf{x})$, P$\phi(t,\mathbf{x})$P${}^{-1}=\phi(t,-\mathbf{x})$, and 
C$\phi(t,\mathbf{x})$C${}^{-1}=\phi^\dag(t,\mathbf{x})$. Furthermore, the action is manifestly invariant under CPT and $\kappa$-deformed Lorentz transformations.


To compute the charges, we first derived the translation charges directly from the Noether theorem, and then we built a covariant phase space approach which was able to reproduce the translation charges, allowing us to compute the remaining ones. As an example, the boost charges are
\begin{align*}
	N_i
	=
	- \frac{1}{2}
	\int d^3p 
	\left\{
	S(\omega_p) \left[
	\frac{\partial a_\mathbf{p}^\dag}{\partial S(\mathbf{p})^i}
	a_\mathbf{p}
	-
	a_\mathbf{p}^\dag \frac{\partial a_\mathbf{p}}{\partial S(\mathbf{p})^i}
	\right]
	+
	\omega_p
	\left[
	b_{\mathbf{p}}
	\frac{\partial b_{\mathbf{p}}^\dag}{\partial \mathbf{p}^i}
	-
	\frac{\partial b_{\mathbf{p}}}{\partial \mathbf{p}^i}
	b_{\mathbf{p}}^\dag
	\right]
	\right\}.
\end{align*}
Furthermore, one can show that $[a, a^\dag] = [b, b^\dag]=1$, which then lets us verify that the deformed charges satisfy the canonical Poincar\'e algebra. One can also build the conjugation operator charge explicitly 
\begin{align}
	\mathcal{C} =
	\int \, d^3p
	\,
	(
	b^\dag_{\mathbf{p}}
	a_\mathbf{p}
	+
	a^\dag_{\mathbf{p}}
	b_{\mathbf{p}}
	).
\end{align}
The creation/annihilation operators algebra then allows us to verify that $[N_i, \text{C}]\neq 0$. More explicitly, one finds that $[N_i, \text{C}]$ is given by
\begin{align}
	&
	\frac{i}{2}
	\int  d^3p 
	\Bigg\{
	S(\omega_p)
	\left[
	\frac{\partial a_{\mathbf{p}}}{\partial S(\mathbf{p})^i}
	b^\dag_{\mathbf{p}}
	-
	a_{\mathbf{p}}
	\frac{\partial b^\dag_{\mathbf{p}} }{\partial S(\mathbf{p})^i}
	+
	\frac{\partial a_{\mathbf{p}}^\dag}{\partial S(\mathbf{p})^i}
	b_{\mathbf{p}}
	-
	a_{\mathbf{p}}^\dag
	\frac{\partial b _{\mathbf{p}}}{\partial S(\mathbf{p})^i}
	\right] + \nonumber \\
	&\qquad \qquad +
	\omega_p
	\left[
	\frac{\partial b_{\mathbf{p}}^\dag}{\partial \mathbf{p}^i}
	a_\mathbf{p}
	-
	b_{\mathbf{p}}^\dag
	\frac{\partial a_\mathbf{p}}{\partial \mathbf{p}^i}
	+
	\frac{\partial b_{\mathbf{p}}}{\partial \mathbf{p}^i}
	a^\dag_{\mathbf{p}}
	-
	b_{\mathbf{p}}
	\frac{\partial a_\mathbf{p}^\dag}{\partial \mathbf{p}^i}
	\right]
	\Bigg\}
\end{align}
Clearly, in the limit $\kappa\rightarrow\infty$ one recovers the canonical $[N_i, \text{C}]=0$. The fact that $[N_i, \text{C}]\neq 0$ has several striking phenomenological consequences. 

In particular, defining $|\mathbf{p}\rangle_a = a_\mathbf{p}^\dag |0\rangle$ and $|\mathbf{p}\rangle_b = b_\mathbf{p}^\dag |0\rangle$, if we have a particle and an antiparticle at rest, then $P_i |\mathbf{p}\rangle_a = P_i |\mathbf{p}\rangle_b = 0$, and $P_0 |\mathbf{p}\rangle_a = M |\mathbf{p}\rangle_b$, $P_0 |\mathbf{p}\rangle_b = M |\mathbf{p}\rangle_b$ (showing that particles and antiparticles have the same mass). Boosting, e.g., in direction $1$, one goes from $|p_1, p_2, p_3\rangle $ to $|\cosh \xi p_1 + \sinh \xi \omega_p, p_2,p_3\rangle$ where $\xi$ is the rapidity. Then one can show that $P_1 |M\sinh\xi,0,0\rangle_a = -S(M\sinh\xi)|M\sinh\xi,0,0\rangle_a$, $P_1 |M\sinh\xi,0,0\rangle_b = M\sinh\xi|M\sinh\xi,0,0\rangle_b$, and analogously for $P_0$. Notice that $\text{C}|\mathbf{p}\rangle_b = |\mathbf{p}\rangle_a$ and viceversa. The difference between particles and antiparticles also reflects itself in a different Lorentz boost for each of them, respectively $-S(E)/M$ and $E/M$. Hence, in a boosted frame, there is a slight difference between the decay probability density function $\mathcal{P}$ between particles and antiparticles. One gets at first order in $1/\kappa$
\begin{eqnarray}
	{\cal P}_{\mbox{\scriptsize part}}(t) & = & \frac{\Gamma E}M\exp \,\left(-\Gamma \,\frac {E}{M}\, t \right),  \label{decay}\\
	{\cal P}_{\mbox{\scriptsize apart}}(t) 
	& = & \Gamma\left(\frac EM - \frac{\mathbf p^2}{\kappa M}\right) \,\exp\,\left [-\Gamma \, \left(\frac EM - \frac{\mathbf p^2}{\kappa M}\right)\, t\,\right ].\label{decaya}
\end{eqnarray}
where $\Gamma=1/\tau$. Since the effects of deformation are given by the factor $\mathbf{p}^2/(\kappa M)$, experimental signatures might be easier to see in a $\mu^+ \mu^-$ pair, since muons are the lightest particles with the best measured lifetimes. 

As a final comment, we note that Greenberg's theorem is not valid in the context of our framework. 



\section*{Acknowledgments}
Parts of these works were supported by funds provided by
the Polish National Science Center, the project number 2019/33/B/ST2/00050.

\end{document}